\documentclass[english,aps,prl,superscriptaddress,twocolumn,showpacs,footinbib]{revtex4-1}
\usepackage{graphicx}  
\usepackage{dcolumn}   
\usepackage{bm}        
\usepackage{amssymb}   
\usepackage{babel}
\usepackage{verbatim}
\usepackage{amsmath}
\usepackage{setspace}
\usepackage{epstopdf}
\graphicspath{{figures/}}
\usepackage{siunitx}
\usepackage{mathtools}

\begin{document}

\widetext

\title{Characterizing spin-bath parameters using conventional and time-asymmetric Hahn-echo sequences}

\author{D. Farfurnik}
\affiliation{Dept. of Applied Physics, Rachel and Selim School of Engineering and Racah Institute of Physics, The Hebrew University of Jerusalem, Jerusalem 9190401, Israel}

\author{N. Bar-Gill}
\affiliation{Dept. of Applied Physics, Rachel and Selim School of Engineering and Racah Institute of Physics, The Hebrew University of Jerusalem, Jerusalem 9190401, Israel}

\date{\today}

\begin{abstract}
Spin-bath noise characterization, which is typically performed by multi-pulse control sequences, is essential for understanding most spin dynamics in the solid-state. Here, we theoretically propose a method for extracting the characteristic parameters of a noise source with a known spectrum, using modified Hahn-echo pulses. By varying the application time of the pulse, measuring the coherence curves of an addressable spin, and fitting these curves to a theoretical function derived by us, we  extract parameters characterizing the physical nature of the noise. Assuming a Lorentzian noise spectrum, we illustrate this method for extracting the correlation time of a bath of nitrogen paramagnetic impurities in diamond, and its coupling strength to the addressable spin of a nitrogen-vacancy center.  First, we demonstrate that fitting conventional Hahn-echo measurements to the explicit coherence function is essential for extracting the correct parameters in the general physical regime, for which common methods relying on the assumption of a slow bath are inaccurate. Second, considering a realistic experimental scenario with a $5\%$ noise floor, we simulate the extraction of these parameters utilizing the asymmetric Hahn-echo scheme. The scheme is effective for samples having a natural homogeneous coherence time ($T_2$) up to two orders of magnitude greater than the inhomogeneous coherence time ($T_2^*$). In the presence of realistic technical drifts for which averaging capabilities are limited, we simulate more than a factor of 3 improvement of the extracted parameter uncertainties over conventional Hahn-echo measurements. Beyond its potential for reducing experiment times by an order-of-magnitude, such single-pulse noise characterization could minimize the effects of long time-scale drifts and accumulating pulse imperfections and numerical errors.   
\end{abstract}
\maketitle

\section{I. Introduction}
\paragraph{}
The quantum dynamics of spins in solid-state systems are often affected by interactions with environmental modes. At low enough temperatures, in which phononic effects become negligible, interactions with a large ensemble of spins, namely the ``spin-bath" in the lattice, may dominate spin dynamics. In systems containing an addressable spin of interest (``the spin qubit"), such as phosphorus donors in silicon and nitrogen-vacancy (NV) centers in diamond, such spin-bath noise acts as the main decoherence source of the system \cite{Tyryshkin2012,Jarmola2012,BarGill2013,Farfurnik2015,DeLange2010}. The first step towards enabling full coherent control of such spin qubits, whose applications range from quantum information processing \cite{Jelezko2006,Bernien2013,Tsukanov2013,Hensen2015} to metrology \cite{Taylor2008,maze2008,Balasubramanian2008,Dolde2011,Barry2016,Chatzidrosos2017}, involves full spectral analysis of the spin-bath noise. 

\paragraph{}
When the spin qubit undergoes a free evolution time $\tau$, which is short compared to the characteristic spin-spin interaction time of the bath $\tau_c$, the application of a resonant $(\pi)$ microwave (MW) pulse, namely a ``Hahn-echo" pulse \cite{Hahn1950}, decouples the spin from static and slow noise terms, thereby increasing its coherence time up to a timescale $T_2$. By applying many such $(\pi)$-pulses, a process often referred to as pulsed ``dynamical decoupling" (DD) \cite{Meiboom1958,Gullion1990,Khodjasteh2005,Ryan2010,Souza2011}, this timescale is significantly extended due to the decoupling from high frequency terms. In recent years, the utilization of such control sequences resulted in orders-of-magnitude enhancements of the coherence properties of superconducting qubits \cite{Cywinski2008,Pokharel2018}, phosphorus donors in silicon \cite{Tyryshkin2012,Wang2012a}, and NV centers \cite{BarGill2013,Farfurnik2015}, as well as enhanced sensitivities of AC magnetometers \cite{Pham2012,Farfurnik2018}.  Beyond enhancing coherence times, DD sequences form an effective tool for the spectral characterization of the spin-bath,  namely its ``spectral decomposition" \cite{Cywinski2008,deSousa2009,Gomez2018}.  Using such techniques, the noise spectrum $S(\omega)$ can be extracted by applying DD sequences with varying numbers of pulses N, and performing deconvolution of the resulting coherence curve. However, the long coherence times associated with the implementation of such schemes result in very long experiments (depending on the signal-to-noise ratio), and increase the system's vulnerability to long-term drifts and accumulation of pulse imperfections \cite{BarGill2012,Romach2019}. Furthermore, the complex deconvolution algorithm procedures introduce additional uncertainties in the resulting noise spectrum \cite{BarGill2012,Romach2019}. In this work, we theoretically propose an alternative method for noise characterization, by time varying a single Hahn-echo pulse, and a simple least-square fitting. While extending previous work focused on the slow noise regime \cite{DeLange2010} to a general spin-bath environment,  we simulate the effectiveness of the method for a typical scenario of NV centers in diamond.  
\section{II. Theoretical Framework}
\subsection{a. Spectral distribution and coherence curves}
\paragraph{}
We begin by introducing the Hamiltonian of a spin qubit under coupling to a spin-bath environment. In the weak coupling regime between the qubit and the spin-bath, the Hamiltonian is given (with the notation of $\hbar \equiv 1$) by \cite{deSousa2009}
\begin{equation}
H(t) =\frac{1}{2}[(\Omega+\eta(t)]\sigma_z,
\end{equation}
where $\Omega$ represents the energy level splitting of the qubit, $\sigma_z$ is the Pauli matrix of the qubit along the $z$ axis and $\eta(t)$ is a classical random variable  representing the coupling to the spin-bath. We define the correlation function of $\eta(t)$ by averaging over noise realizations 
\begin{equation}
s(t-t')=\langle \eta(t)\eta(t') \rangle,
\end{equation} and its corresponding spectral distribution by the integral 
\begin{equation}
S(\omega)=\int_{-\infty}^{\infty}e^{i\omega t}s(t)dt.
\end{equation}
\paragraph{}
Assuming that the spin qubit is initialized to a certain state, we apply a control sequence characterized by a time-domain function $f(t)$, at the resonant MW frequency determined by its energy levels. This function represents the effect of the pulse sequence on the spin state as a function of time (e.g. equals 1 for no change in the Bloch vector, -1 for a spin-flip). The ``coherence function", describing the fidelity between the initial state and the state at time $T$, is given by \cite{Cywinski2008,deSousa2009}
\begin{equation}
W(T)=\left | \left\langle \exp\left({-i\int_0^T \eta(t')f(t')dt'}\right) \right\rangle \right | \label{eq:Wtime}.
\end{equation}
If $\eta(t)$ has Gaussian statistics, only the second order correlation $s(t)$ contributes to \eqref{eq:Wtime} \cite{deSousa2009,Zwick2016}, and the coherence function yields 
\begin{equation} 
W(T)=\exp\left({-\int_0^{\infty} \frac{d \omega}{\pi} S(\omega) \frac{F(\omega T)}{\omega^2}}\right) \label{eq:Wfreq},
\end{equation}
where $F(\omega)=\frac{\omega^2}{2} \left |\mathfrak{F}[f(t)]\right |^2$ is proportional to the squared Fourier transform of the time-domain function $f(t)$ and referred to as the ``filter function" of the control sequence \cite{deSousa2009,Witzel2007}. For non-Gaussian statistics, higher-order correlations have to be taken into account \cite{deSousa2009,Zwick2016}.
\paragraph{}
 The spectral distribution of the spin-bath $S(\omega)$ can be extracted experimentally from fitting to eq. \eqref{eq:Wfreq}, by applying various control sequences represented by known filter functions $F(\omega)$, measuring the coherence curve at different times $W(t)$, and performing numerical deconvolution according to \eqref{eq:Wfreq}. The numerically extracted curve can be fitted to various functional forms to obtain the best-fitting spectral distribution and the corresponding physical parameters. This method, which is referred to as ``spectral decomposition" \cite{Cywinski2008,deSousa2009}, has been recently used to characterize the noise surrounding NV centers in diamond with the Carr-Purcell-Meiboom-Gill \cite{BarGill2012} and DYSCO \cite{Romach2019} sequences serving as the control schemes. Beyond the long experiment times (several hours up to days) involved in its experimental realization, such an analysis results in significant vulnerability to experimental drifts, pulse imperfections and numerical inaccuracies. Due to the complexity in taking these imperfections into account in the functional form of $F(\omega)$, the extracted power spectrum typically contains unwanted artifacts. Furthermore, the complex deconvolution process introduces additional uncertainties \cite{BarGill2012,Romach2019}.  
 \subsection{b. Asymmetric Hahn-echo analysis}
\begin{figure}[!t]	
	\includegraphics[width=1\columnwidth]{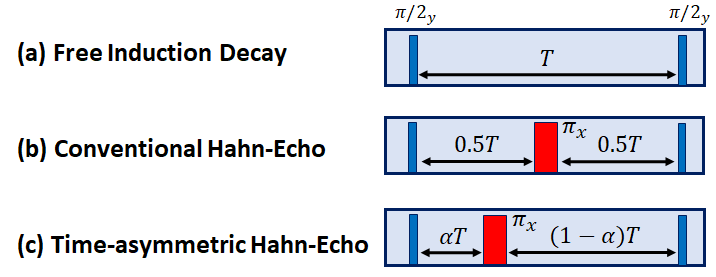}
	\caption{(Color online) Conventional free induction decay and Hahn-echo sequences, and the time-asymmetric Hahn-echo sequence.}
	\label{fig:schemes}
\end{figure}

\paragraph{}
We propose an alternative method for spectral noise characterization, using a single Hahn-echo MW ($\pi$)-pulse \cite{Hahn1950}. The ($\pi$)-pulse is applied during free spin evolution, at an intermediate time $\tau=\alpha T$, with $T$ the total experiment time and $0 \le \alpha \le 1$. The limits of $\alpha=0,1$ correspond to free induction decay (FID), and the case of $\alpha=\frac{1}{2}$ yields conventional Hahn-echo (Fig. \ref{fig:schemes}). 
A single experiment involves sweeping the total time $T$, while keeping $\alpha$ constant, and extracting the coherence curve as a function of time similarly to conventional Hahn-echo measurements. While standard spectral decomposition techniques incorporate hundreds of pulses, the duration of a single-pulse asymmetric Hahn-echo experiment is no longer than a conventional Hahn-echo experiment, and it does not suffer from pulse imperfections. 
Considering a well-known functional form for the noise spectrum, we calculate the functional form of the coherence curve for different values of $\alpha$, $W(T,\alpha)$  from Eq. \eqref{eq:Wfreq}. For a given value of $\alpha$, the resulting decay incorporates both spin refocusing due to the $(\pi)$-pulse at time $2\alpha T$, and the remaining $T(1-2\alpha)$ time of FID dynamics. 
The dependence of the various decay curves on $\alpha$ is dictated uniquely by the nature of the spin-bath noise, enabling the extraction of its physical parameters by standard least-square fitting to an analytical function. In order for the variation in $\alpha$ to provide additional information on the noise over conventional Hahn-echo measurements, both relative FID and Hahn-echo parts have to significantly contribute to the dynamics \cite{Zwick2016,Mouro2016}. Considering a reasonable temporal resolution of two nanoseconds, applying $(\pi)$-pulses is required at times two orders of magnitude shorter than the total experiment time. This translates to the requirement on the natural FID and Hahn-echo coherence times $T_2(\alpha=0.5)/T_2^*< 100$.
 \subsection{c. General Hahn-echo coherence function under a Lorentzian bath}   
\paragraph{}
A well-known functional form for describing the noise spectrum represented by a spin-bath in solids involves Gaussian statistics $\eta(t)$, and a Lorentzian noise spectrum \cite{Uhlenbeck1930,deSousa2009,BarGill2012}
\begin{equation}
S(\omega)=\frac{b^2 \tau_c}{\pi} \frac{1}{(\omega \tau_c)^2+1},
\end{equation}
where $\tau_c$ is the correlation time of the bath and $b$ is the coupling strength to the spin qubit of interest. The spin-bath correlation function in this case is given by:
\begin{equation}
s(t)=\langle \eta(t) \eta(0) \rangle=b^2 \exp \left ( -\frac{|t|}{\tau_c} \right ).
\label{eq:correlation}
\end{equation}
When the duration of the $(\pi)$-pulse is much shorter than the experiment time $T$, the pulse can be described as an instantaneous rotation of the spin qubit at $\tau=\alpha T$. Such a rotation flips the direction in which the spin accumulates phase, and can be represented by a time-domain control of the form 
\begin{equation}
f(t) =
\begin{cases}
1 &  0\le t\le \alpha T  \\
-1       &  \alpha T \le t \le T.
\end{cases}
\end{equation}
As a result, according to Eq. \eqref{eq:Wtime} under the assumption of Gaussian distribution of $\eta(t)$, the coherence function is given by
\begin{align}
\label{eq:exact}
&W(T,\alpha)=\left | \left\langle \exp\left[-i \left( \int_0^{\alpha T}-\int_{\alpha T}^T \right) \eta(t)dt \right ] \right\rangle \right | \nonumber \\
&=\exp\left[\left( \int_0^{\alpha T}-\int_{\alpha T}^T \right)dt \int_0^{t}s(t')dt' \right ] \nonumber \\
&=\exp\left [-b^2 \tau_c^2 \left (\frac{T}{\tau_c}-3+2e^{-\frac{\alpha T}{\tau_c}}+2e^{-\frac{(1-\alpha)T}{\tau_c}}-e^{-\frac{T}{\tau_c}}\right ) \right ] , 
\end{align}
For any single experiment with a given value of $\alpha$, a simple least-square fitting of the measured coherence curve to Eq. \eqref{eq:exact} can be used to evaluate the realistic noise parameters $b$,$\tau_c$. \\

\section{III. Simulations}
\subsection{a. General regime versus the assumption of a slow noise}
\begin{figure}[!t]	
	\includegraphics[width=1\columnwidth]{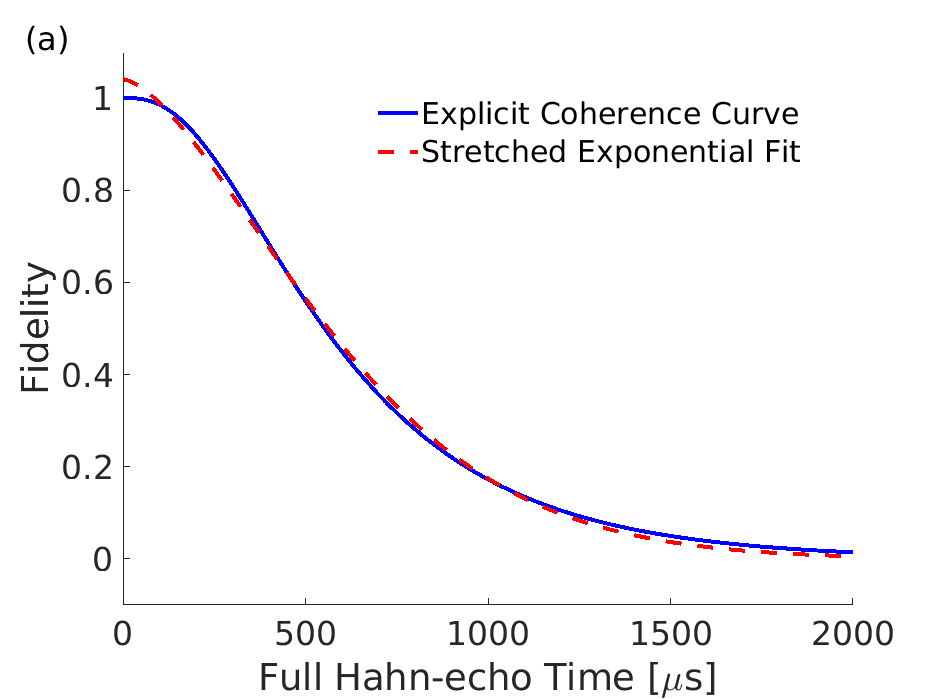}
	\includegraphics[width=1\columnwidth]{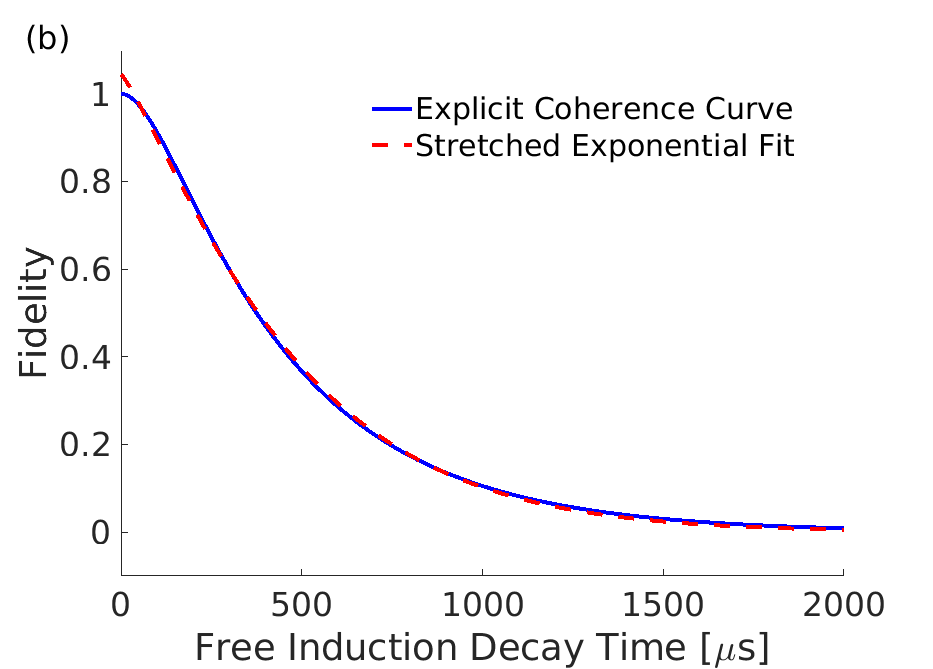}
	\caption{(Color online) Coherence curves as a function of time generated from eq. \eqref{eq:exact} considering a spin-bath with $\tau_c=100$ $\SI{}{\micro\second}$,  $b=5$ $\SI{}{\kilo\hertz}$ and no typical experimental readout errors (blue, solids), and their corresponding fits to stretched exponential decay functions (red, dashed) for conventional (a) Hahn-echo ($\alpha=\frac{1}{2}$) and (b) FID ($\alpha=0$) experiments.}
	\label{fig:SlowVsExplicit}
\end{figure}

Spin-bath parameters are often extracted from conventional Hahn-echo and FID measurements, under the assumption of a `slow noise regime', $T \ll \tau_c$ \cite{deSousa2009,DeLange2010}. Adopting such an approach, the Hahn-echo coherence curve ($\alpha=\frac{1}{2}$) yields $\exp{[-(T/T_2)^3]}$, with the characteristic decay time $T_2=(\frac{12\tau_c}{b^2})^{1/3}$, and the FID coherence curve ($\alpha=0$) yields $\exp{[-(T/T_2^*)^2]}$, with the characteristic decay time $T_2^*=\frac{\sqrt{2}}{b}$. Consequently, the spin-bath parameters are typically extracted from fitting these curves to stretched exponential functions of the form $\exp[-(T/T_2^{(*)})^p]$, followed by plugging the extracted coherence times into the analytical relations $b=\frac{\sqrt{2}}{T_2^*}$, $\tau_c=\frac{T_2^3b^2}{12}$. In the general physical regime for which the assumption of slow noise is not valid, however, such an analysis may lead to an inaccurate evaluation.
\paragraph{}
We now highlight the importance of extracting the spin-bath parameters in the general physical regime using the explicit expression \eqref{eq:exact}, by simulating a realistic scenario of a Lorentzian noise relevant in the $\sim$ ppm nitrogen regime, with $\tau_c=100$ $\SI{}{\micro\second}$ and $b=5$ $\SI{}{\kilo\hertz}$ \cite{BarGill2012,Farfurnik2015} (Fig \ref{fig:SlowVsExplicit}). In the ideal case excluding any measurement noise,  while fitting the Hahn-echo and FID to the explicit function (blue solid lines in Fig. \ref{fig:SlowVsExplicit}) predicts the correct parameters with zero uncertainty, the common fitting approach under the assumption of slow noise (red, dashed lines in Fig. \ref{fig:SlowVsExplicit}) yields $b=2.88\pm0.46$ $\SI{}{\kilo\hertz}$, $\tau_c=225\pm97$ $\SI{}{\micro\second}$, in a complete disagreement with the correct parameters

\subsection{b. Shot-noise-limited measurements}
\begin{figure}[!t]	
	\includegraphics[width=1\columnwidth]{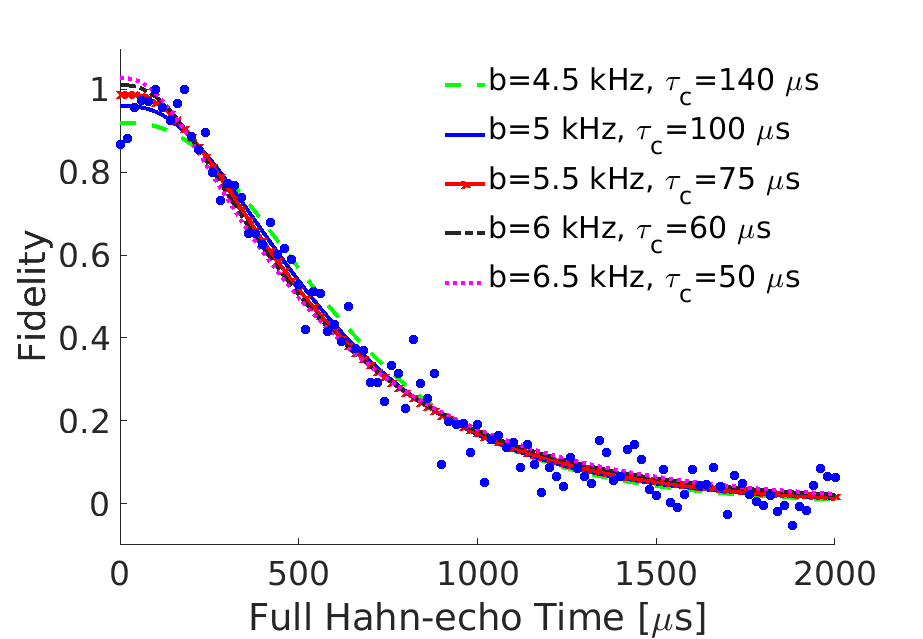}
	\caption{(Color online) Fitting to a conventional Hahn-echo ($\alpha=\frac{1}{2}$) and coherence curve as a function of time. The simulated data (dots) were generated considering a spin-bath with $\tau_c=100$ $\SI{}{\micro\second}$,  $b=5$ $\SI{}{\kilo\hertz}$ and a typical experimental readout error of $5\%$. Different decay curves (lines) fitted to the data using Eq. \eqref{eq:exact} and different spin-bath parameters fit the data within an uncertainty of $\Delta \chi_\nu^2=0.14$.}
	\label{fig:Echo_Different}
\end{figure}

\begin{figure}[!t]	
	\includegraphics[width=0.49\columnwidth]{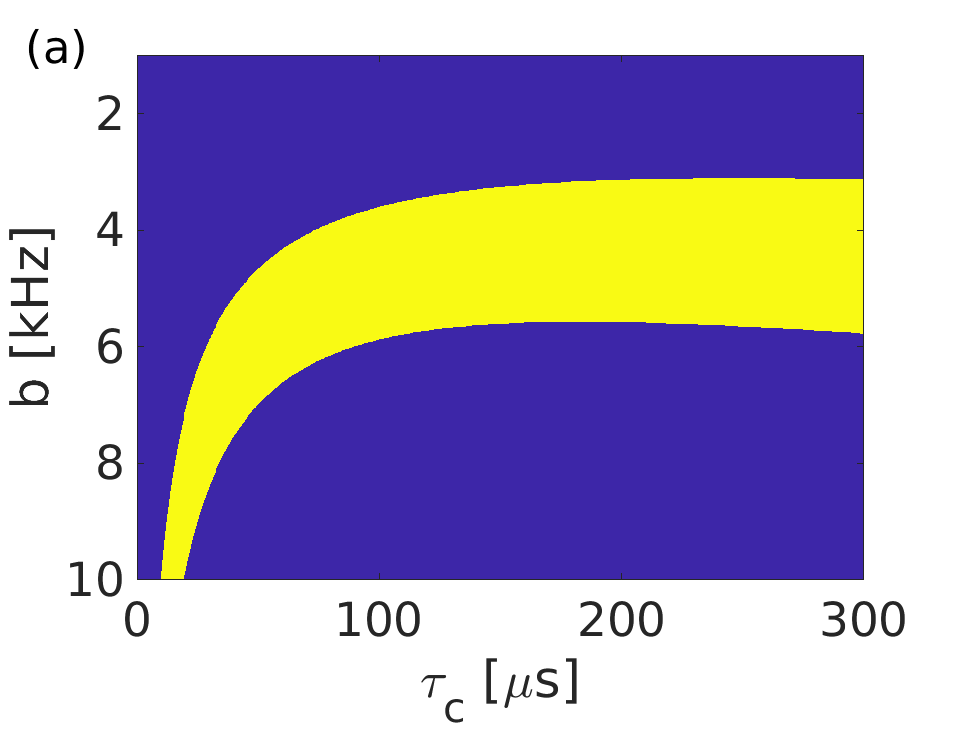}
	\includegraphics[width=0.49\columnwidth]{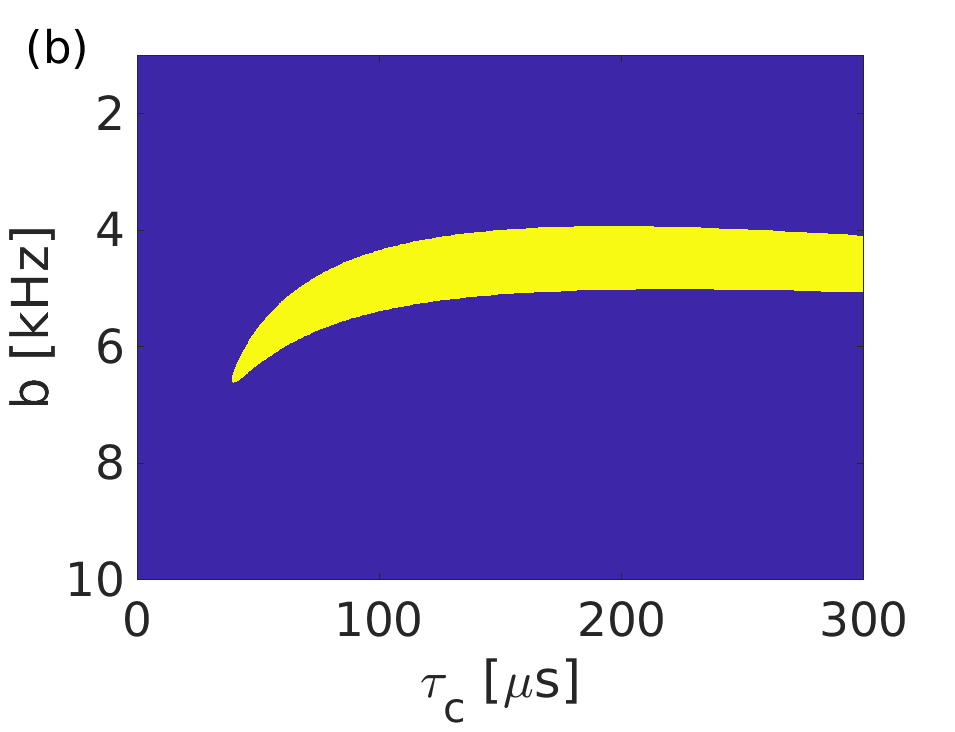}
	\includegraphics[width=0.49\columnwidth]{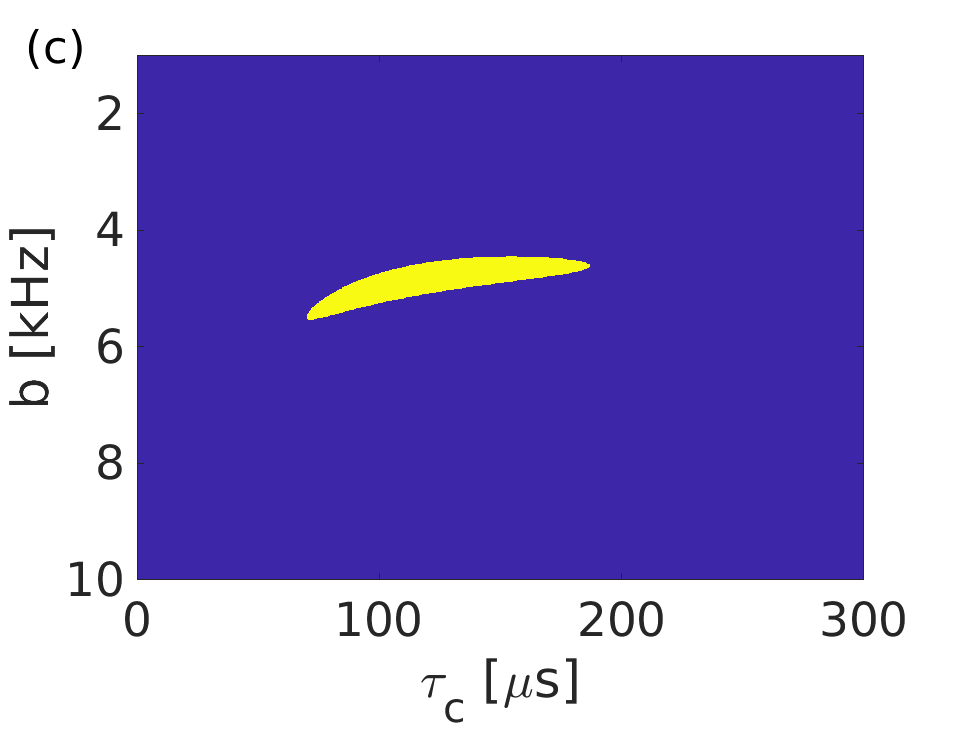}
		\includegraphics[width=0.49\columnwidth]{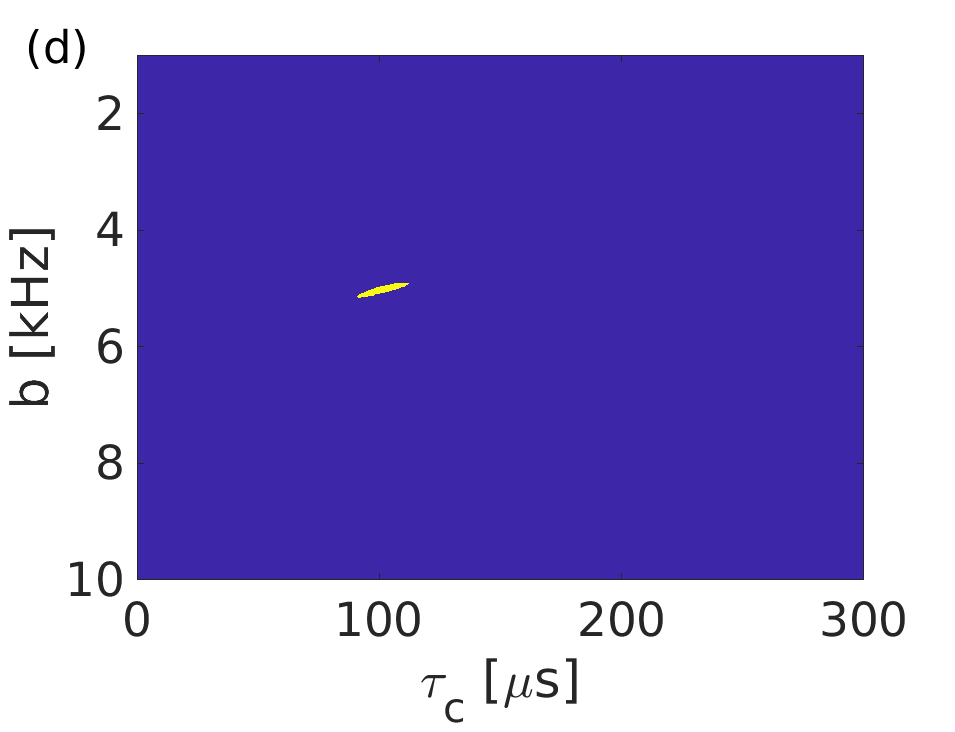}
		\includegraphics[width=0.49\columnwidth]{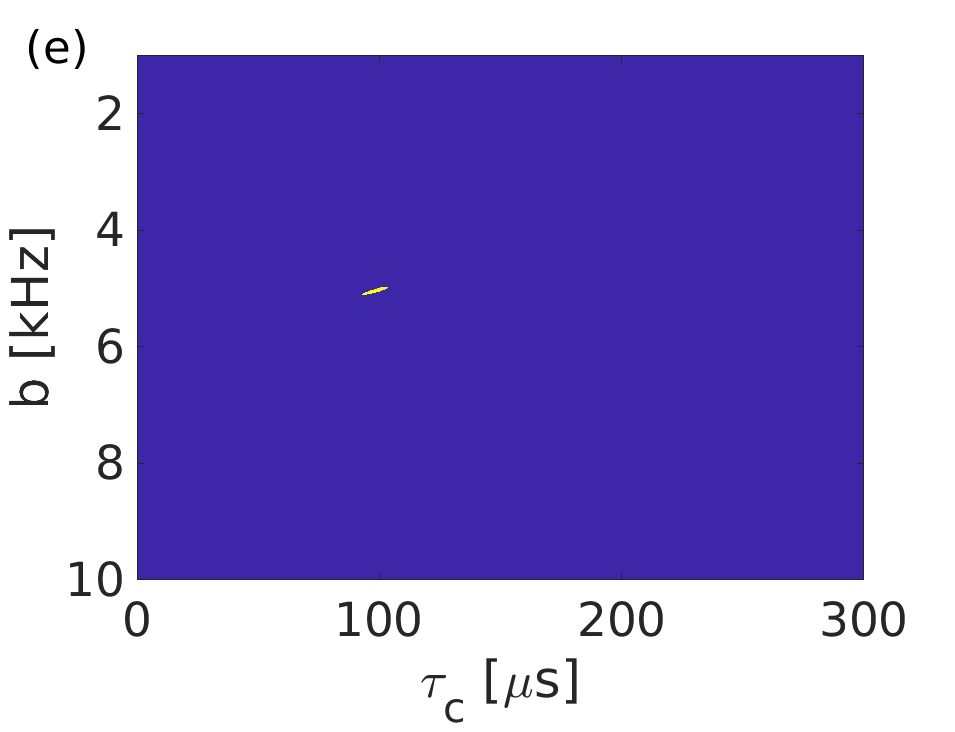}
		\includegraphics[width=0.49\columnwidth]{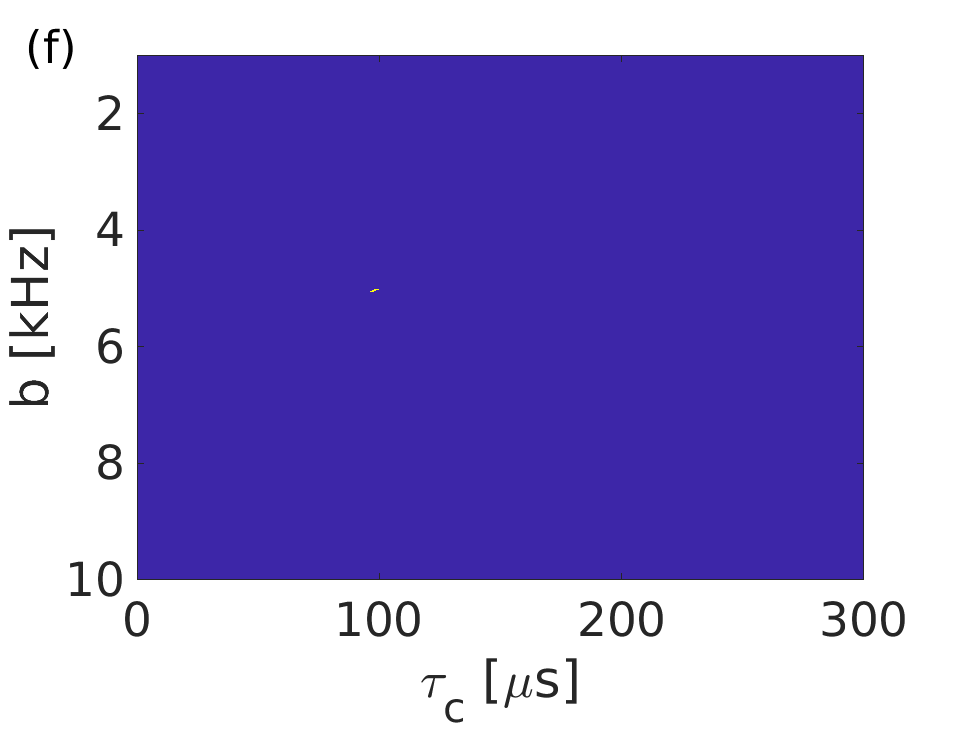}
	\caption{(Color online) Parameter pairs fitting the Hahn-echo ($\alpha=\frac{1}{2}$ ) explicit coherence curve [Eq. \eqref{eq:exact}] for a spin-bath with parameters $\tau_c=100$ $\SI{}{\micro\second}$,  $b=5$ $\SI{}{\kilo\hertz}$ within the $\chi^2_{\nu}$ uncertainty of 0.14 (yellow regions) as a function of the number of averages. Initial shot-noise simulated with a signal-to-noise ratio of 1. (a) 25, (b) 100, (c) 400, (d) 1e4, (e) 4e4, and (f) 2.5e5 averages. The convergence to a single point indicates the unlimited precision achievable by averaging.}
	\label{fig:Average_Echo_Shot}
\end{figure}

\paragraph{}
As opposed to the ideal case described in the previous section, realistic  experiments involve measurement noise, under which even the most accurate fitting can never result in zero uncertainties. The most common measurement noise source related to the photon detection efficiency can be treated as a `shot-noise', whose uncertainty reduces as the square-root of the number of averages. For simulating a conventional Hahn-echo experiment with the spin-bath parameters $\tau_c=100$ $\SI{}{\micro\second}$, $b=5$ $\SI{}{\kilo\hertz}$, we use a Gaussian distribution to randomly generate realistic measurement noise fluctuations of $5\%$ \cite{BarGill2012,Farfurnik2015} around the $\alpha=\frac{1}{2}$ coherence curve explicitly calculated from Eq. \eqref{eq:exact}. Considering these realistic features, the simulated data were fitted with different parameter pairs $\{b,\tau_c\}$ to Eq. \eqref{eq:exact} using a least-square algorithm. Beyond qualitative similarities between fitted curves with different parameter pairs (represented as lines in Fig. \ref{fig:Echo_Different}), they agree quantitatively within the uncertainty of the reduced goodness-of-fit measure $\Delta \chi^2_\nu=0.14$ (for a typical measurement with 100 degrees of freedom). Within this region, the probability of any parameter pair to represent a noise source fitting the resulting coherence curve data is at least $70\%$.
\paragraph{}
A two-dimensional color map, illustrating the set of parameters fitting the Hahn-echo curve within this uncertainty (Fig. \ref{fig:Average_Echo_Shot}), provides evidence of the inverse correlations between $b$ and $\tau_c$. These correlations correspond to the direct (inverse) proportionality of $b$ ($\tau_c$) to the density of the spins in the bath. Due to the nature of the considered shot-noise measurement source, further averaging the Hahn-echo experiment could limitlessly reduce the uncertainties of the extracted parameters  (as illustrated by the convergence to a single point in Fig. \ref{fig:Average_Echo_Shot}).

\subsection{c. Noise floor imposed by technical drifts}
\begin{figure}[!t]	
	\includegraphics[width=0.49\columnwidth]{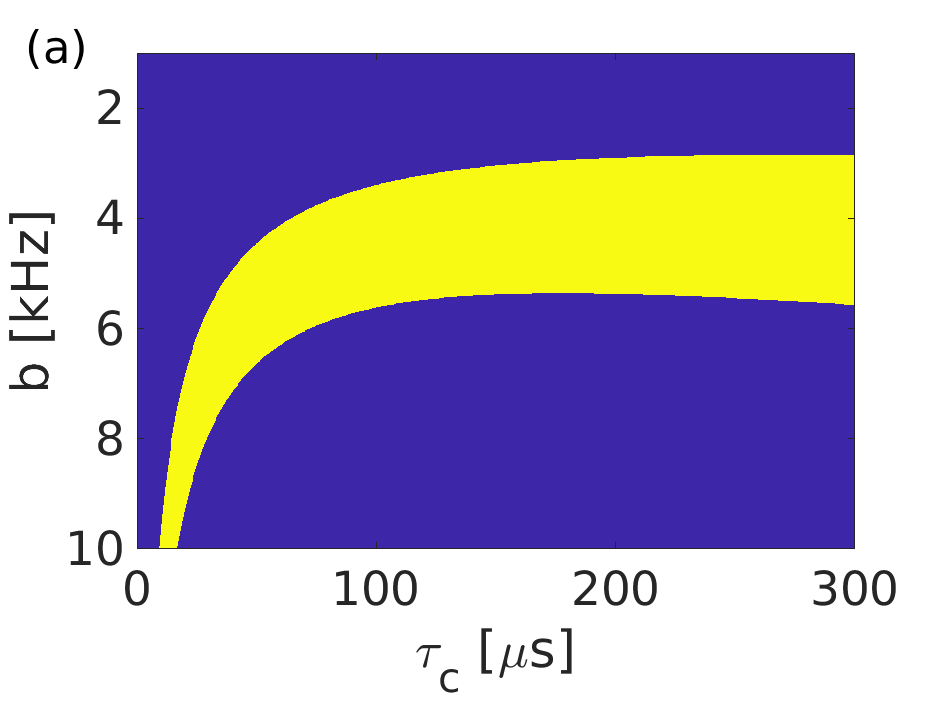}
	\includegraphics[width=0.49\columnwidth]{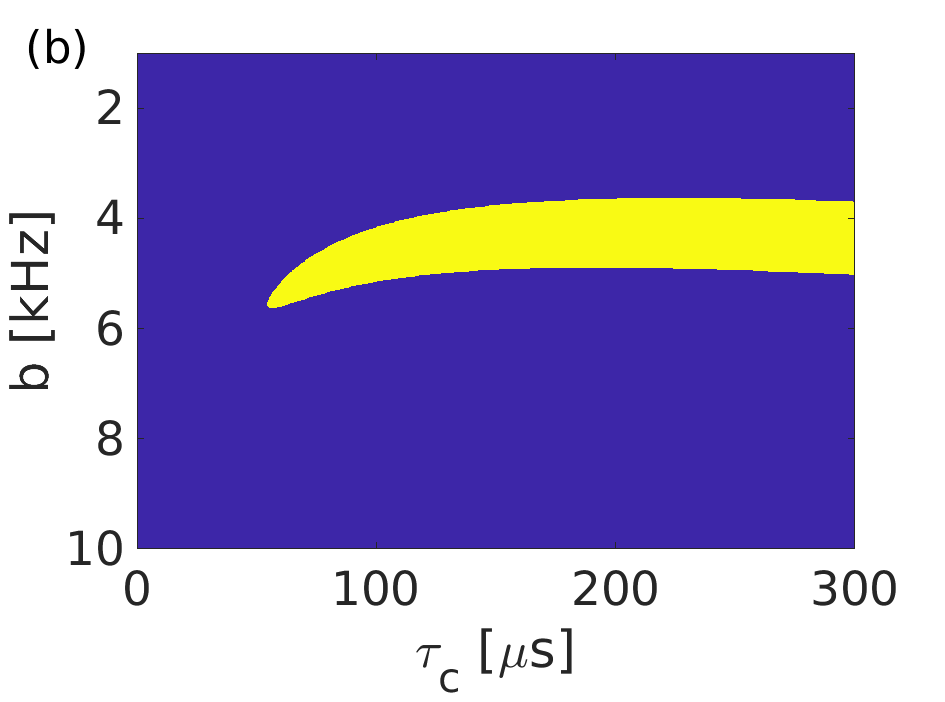}
	\includegraphics[width=0.49\columnwidth]{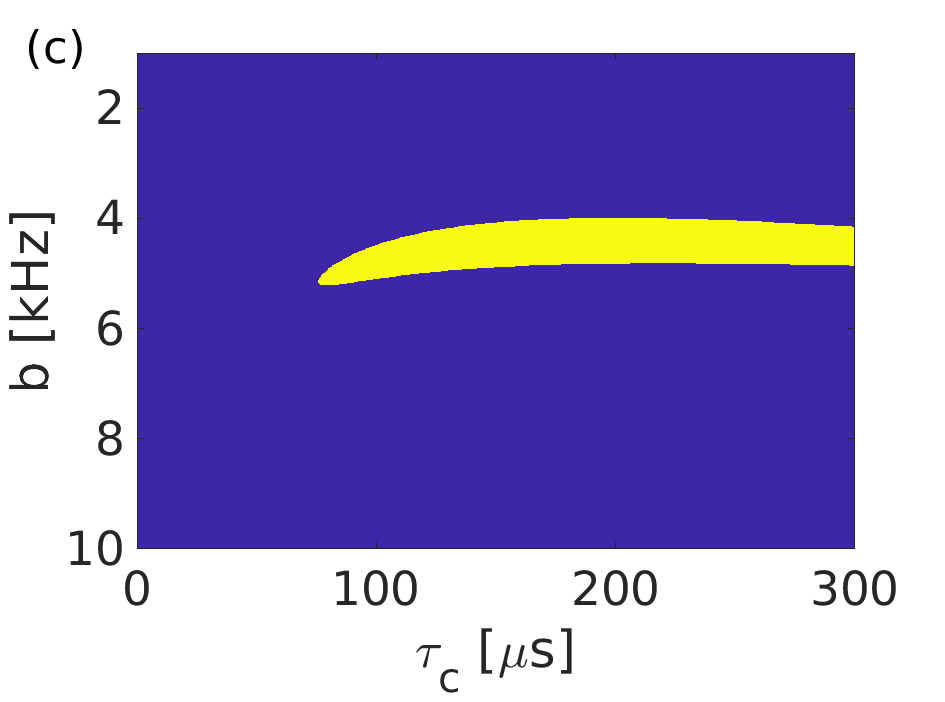}
	\includegraphics[width=0.49\columnwidth]{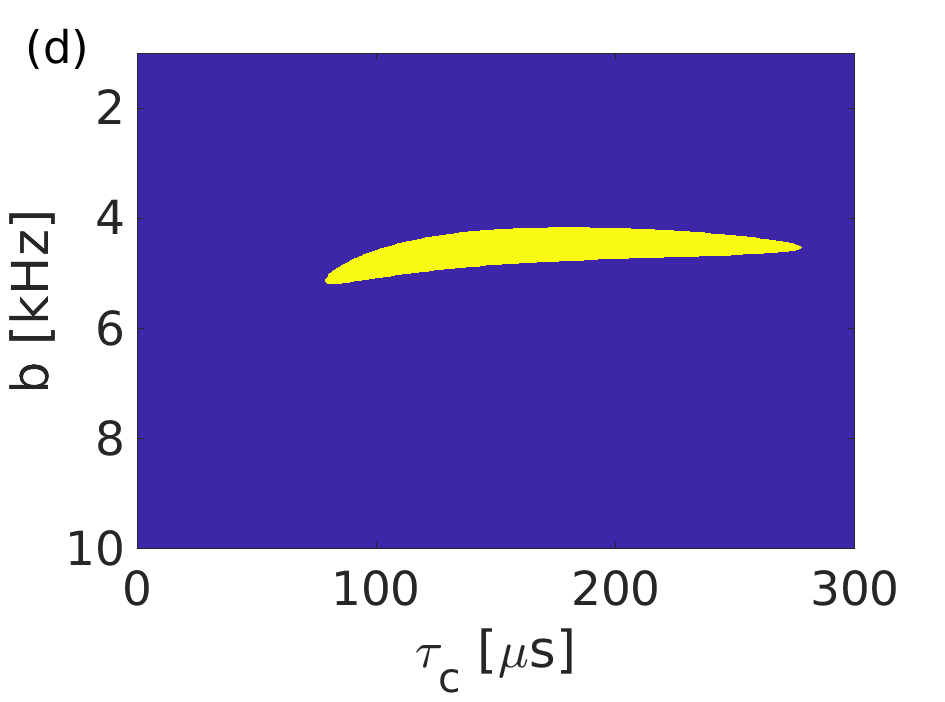}
	\includegraphics[width=0.49\columnwidth]{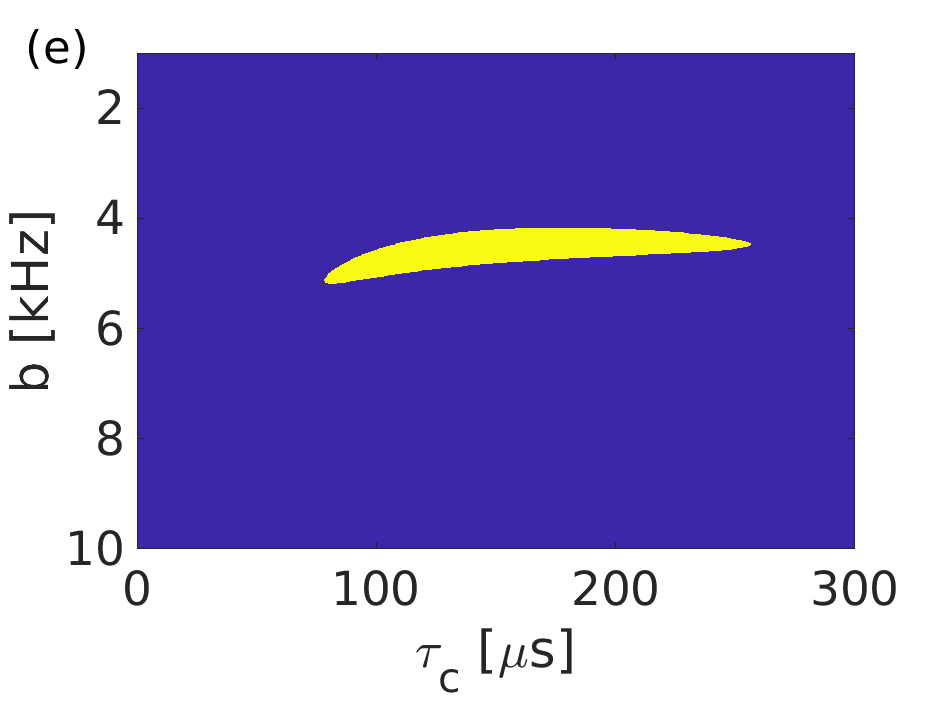}
	\includegraphics[width=0.49\columnwidth]{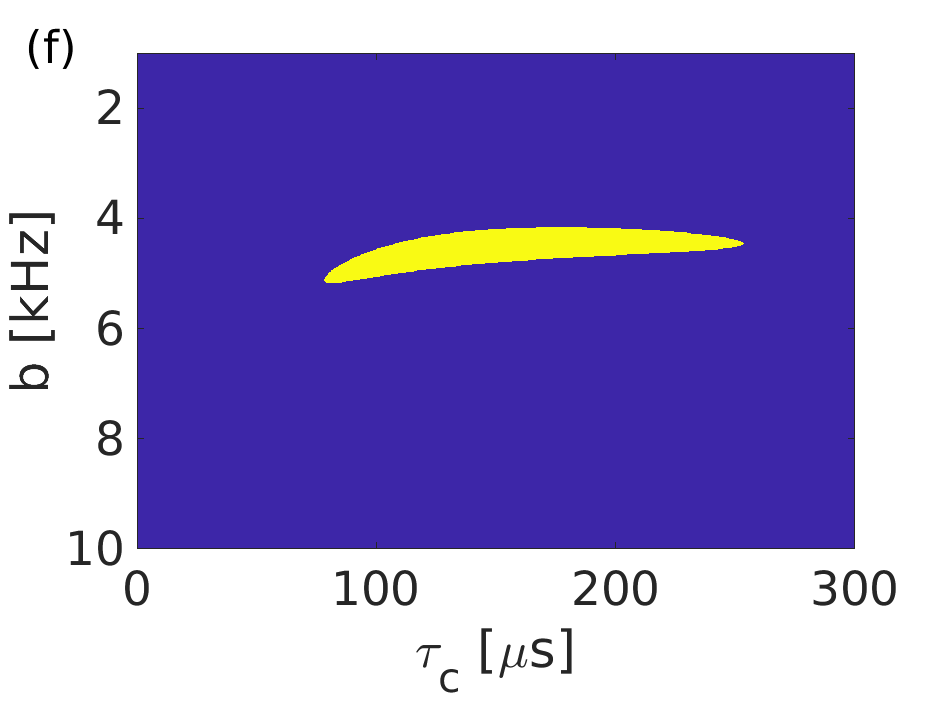}
	\caption{(Color online) Parameter pairs fitting the Hahn-echo ($\alpha=\frac{1}{2}$ ) explicit coherence curve [Eq. \eqref{eq:exact}] for a spin-bath with parameters $\tau_c=100$ $\SI{}{\micro\second}$,  $b=5$ $\SI{}{\kilo\hertz}$ within the $\chi^2_{\nu}$ uncertainty of 0.14 (yellow regions) and a $5\%$ noise floor as a function of the number of averages. Initial shot-noise simulated with a signal-to-noise ratio of 1. (a) 25, (b) 100, (c) 400 (corresponding to a signal-to-noise ratio of $5\%$), (d) 1e4, (e) 4e4, and (f) 2.5e5 averages. Further averaging no longer reduces the extracted parameter uncertainties when the noise floor dominates.}
	\label{fig:Average_Echo_Floor}
\end{figure}

\begin{figure}[!t]	
	\includegraphics[width=0.49\columnwidth]{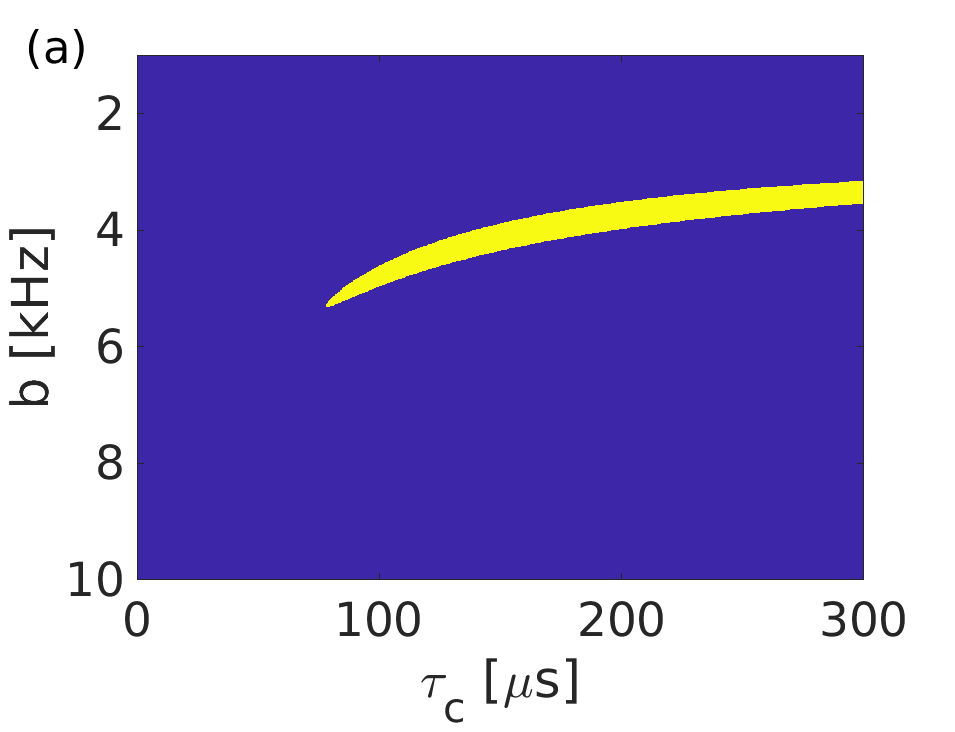}
	\includegraphics[width=0.49\columnwidth]{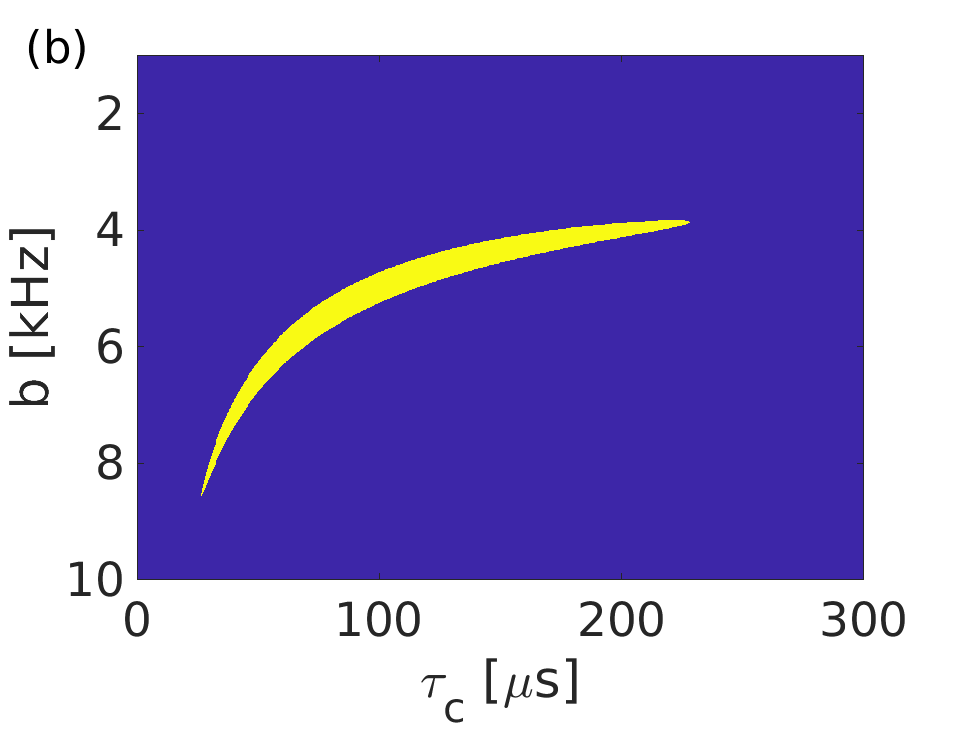}
	\includegraphics[width=0.49\columnwidth]{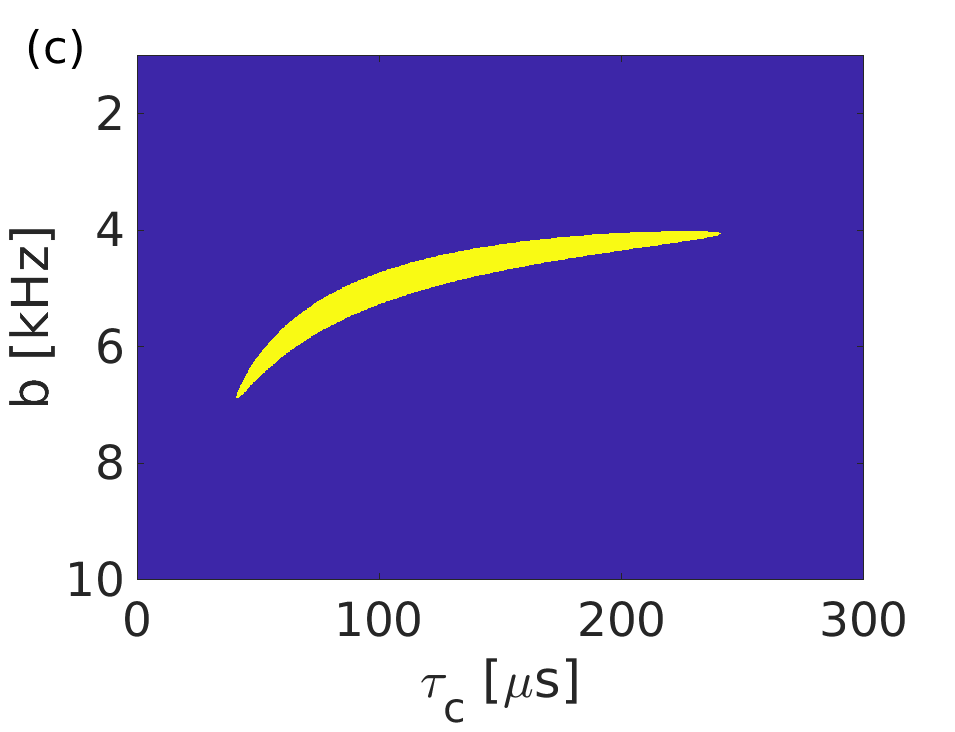}
	\includegraphics[width=0.49\columnwidth]{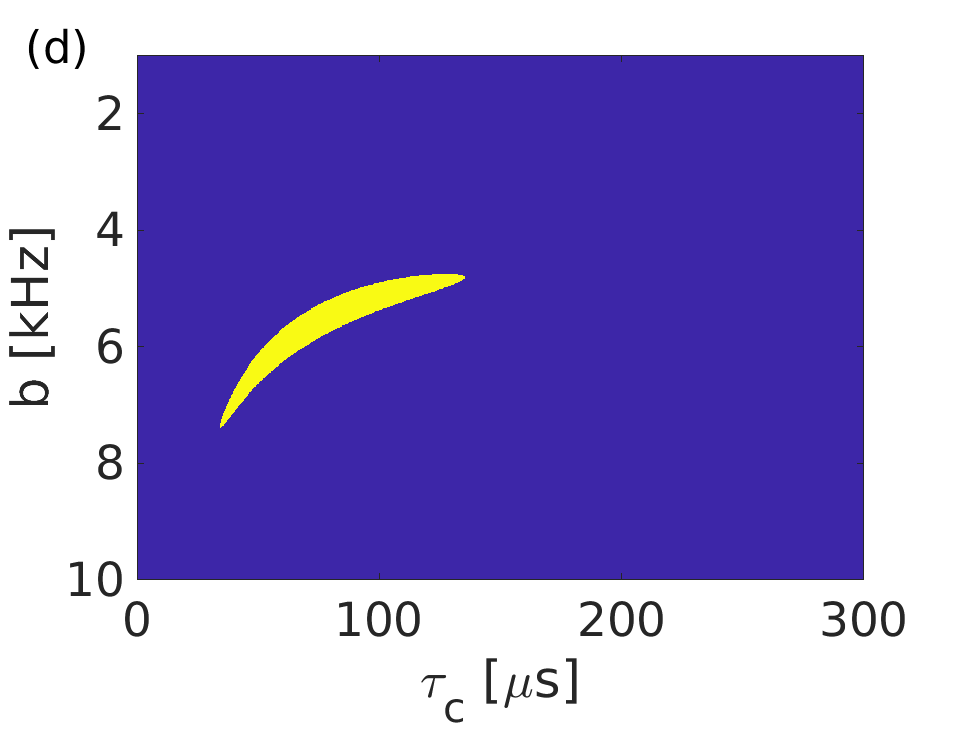}
	\includegraphics[width=0.49\columnwidth]{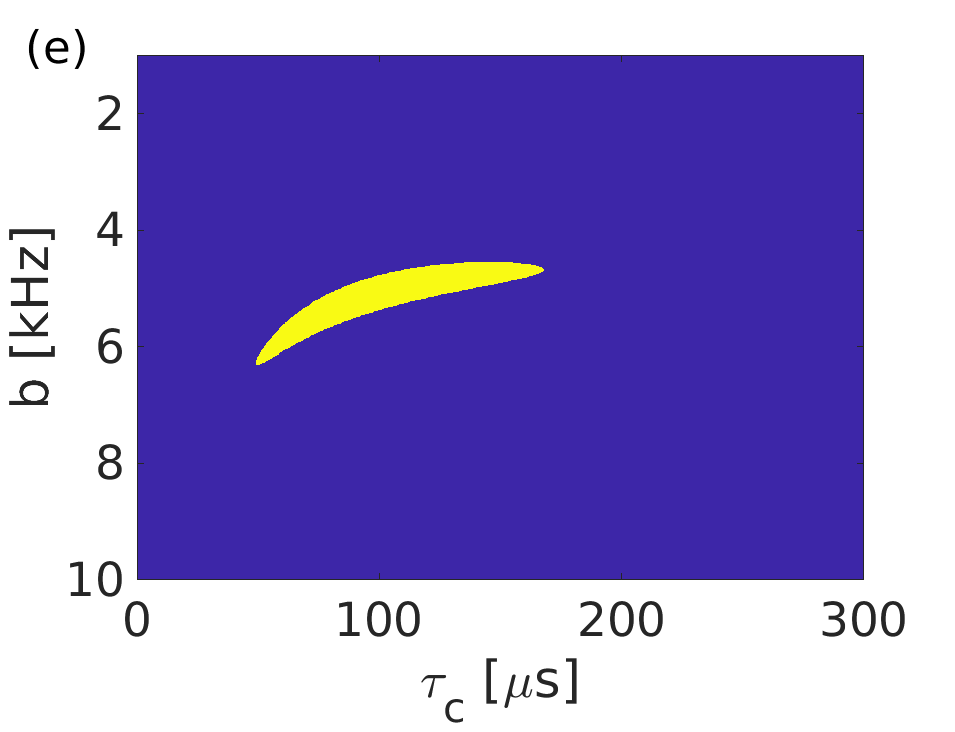}
	\includegraphics[width=0.49\columnwidth]{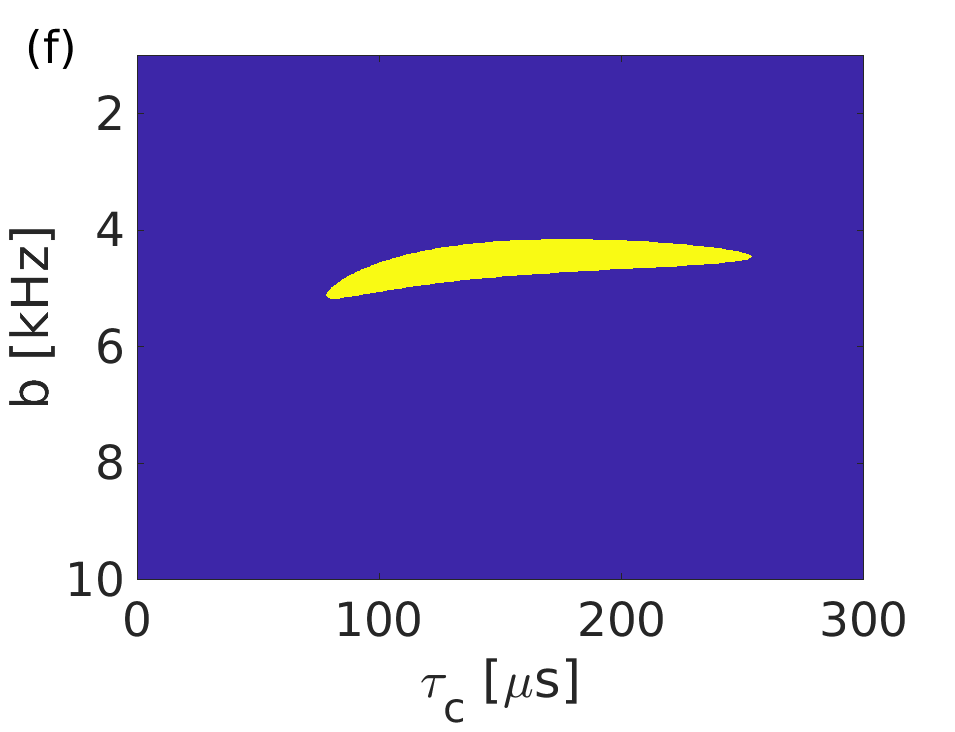}
	\includegraphics[width=1\columnwidth]{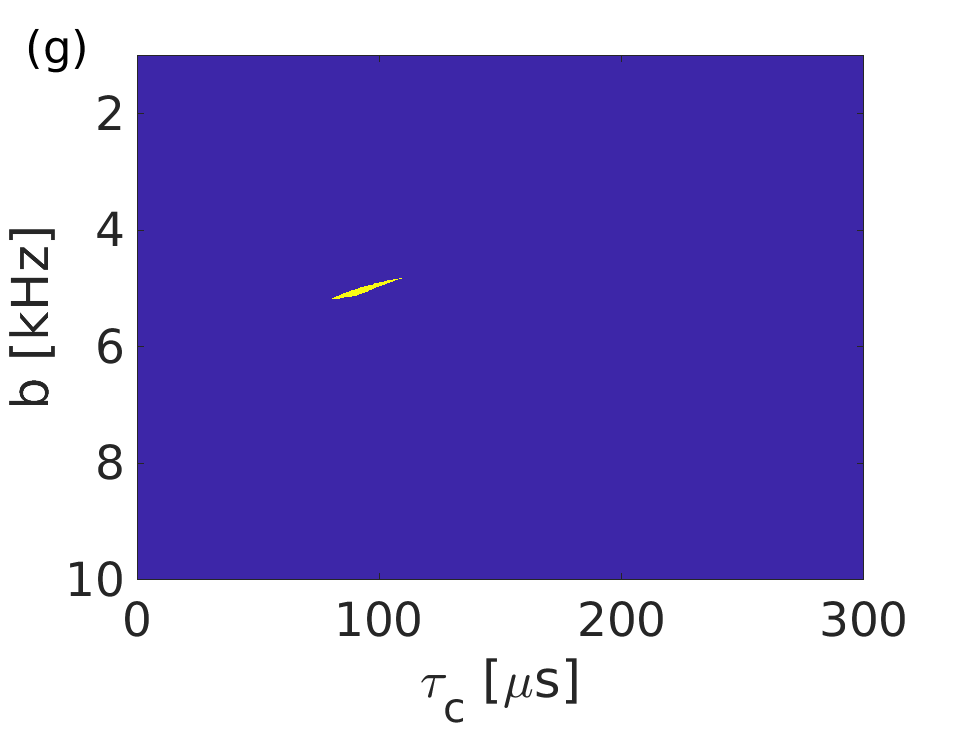}
	\caption{(Color online) Parameter pairs fitting the explicit coherence curves [Eq. \eqref{eq:exact}] of asymmetric Hahn-echo experiments with $\alpha=$ (a) 0, (b) 0.1, (c) 0.2, (d) 0.3, (e) 0.4, (f) 0.5,  for a spin-bath with parameters $\tau_c=100$ $\SI{}{\micro\second}$,  $b=5$ $\SI{}{\kilo\hertz}$ within the $\chi^2_{\nu}$ uncertainty of 0.14 (yellow regions), a $5\%$ noise floor, an initial shot-noise with a signal-to-noise ratio of 1 and 2.5e5 averages. (g) The shared parameter pairs obtained from the intersection of the regions in (a)-(f), and resulting in the extracted parameters $\tau_c=95\pm 15$ $\SI{}{\micro\second}$, $b=5.00 \pm 0.17$ $\SI{}{\kilo\hertz}$.}
	\label{fig:Uneven_Floor}
\end{figure}

\paragraph{}
While the previous section discussed extracting spin-bath parameters for a shot-noise-limited measurement noise, realistic experiments often involve additional drifts in the measured signal, such as laser and modulator instabilities. As opposed to shot-noise sources, such drifts can be proportional to the measured signal itself. Considering an experiment with $N$ detected photons, such realistic drifts can result in a measured signal of $rN$, and a consequent signal-to-noise ratio of $N/\sqrt{N+r^2 N^2}$. Even at infinite number of averages, this signal-to-noise ratio will yield a typical constant noise floor $1/r$, estimated in realistic experiments as a few percent. While conventional Hahn-echo fitting is sufficient for shot-noise-dominated systems, additional time-asymmetric Hahn-echo pulses could significantly reduce the uncertainties of the extracted spin-bath parameters in the presence of a noise floor unmitigable by averaging. For fitting a conventional Hahn-echo coherence curve, the achievable uncertainties of the extracted parameters will saturate at the number of averaging corresponding to the collection of $N$ photons satisfying $1/\sqrt{N}<1/r$. Simulating such an averaging under a spin-bath with the parameters $\tau_c=100$ $\SI{}{\micro\second}$, $b=5$ $\SI{}{\kilo\hertz}$ and a noise floor of $5\%$ (as illustrated in Fig. \ref{fig:Average_Echo_Floor}) leads to the extracted parameters $\tau_c=166\pm 88$ $\SI{}{\micro\second}$, $b=4.7 \pm 0.5$ $\SI{}{\kilo\hertz}$.
\paragraph{}
Reducing the uncertainties of the extracted spin-bath parameters requires additional physical information. The concept of mitigating a constant noise floor was introduced in the field of ``weak measurements" \cite{Hosten2008,Dixon2009}, which involves amplifying the signal while keeping the noise floor constant. Here, we adopt the alternative approach of analyzing data from independent experiments sampling different physical regimes. Additional asymmetric Hahn-echo experiments implementing pulses at times $\alpha\neq 0, \frac{1}{2}$ will sample new physical regimes incorporating different fractions of refocused and free evolution-mediated  dynamics. For each value of $\alpha$, we perform an independent least-square-fitting procedure, to obtain the set of parameter pairs fitting Eq. \eqref{eq:exact} within the uncertainty of the goodness-of-fit measure $\chi^2_{\nu}$. Considering six uneven-echo experiments placed at $\alpha=0,0.1,0.2,0.3,0.4,0.5$ implementable by common pulsing cards having a temporal resolution of two nanoseconds, while the accuracy of a single experiment is limited by the noise floor, filtering the parameter pairs shared for all the curves results in significantly reduced uncertainties (Fig. \ref{fig:Uneven_Floor}). For the experimental conditions described above, the extracted spin-bath parameters, $\tau_c=95\pm 15$ $\SI{}{\micro\second}$, $b=5.00 \pm 0.17$ $\SI{}{\kilo\hertz}$ correctly predict the simulated noise (as opposed to the values $\tau_c\sim 200$ $\SI{}{\micro\second}$, $b\sim 2.7$ $\SI{}{\kilo\hertz}$ predicted by the conventional approach assuming a slow bath), and improve the precision over the saturated uncertainties of conventional Hahn-echo fitting by more than a factor of 3.  
Note that additional experiments at $\alpha>0.5$ sample equivalent physical regimes as their corresponding experiments at $|1-\alpha|$, and thus are not expected to provide a significant added value over the simulations introduced here.
\section{IV. Conclusions}
\paragraph{}
To summarize, we highlight the importance in considering the general physical expression for extracting characteristic spin-bath noise parameters, and propose a novel uneven-echo method for implementing such noise spectroscopy. While the parameters can be extracted from fitting conventional Hahn-echo curves, asymmetric pulses may further mitigate the effects of technical drifts and reduce the uncertainties of the extracted parameters. Although this method does not provide full spectral information, it can offer an order-of-magnitude reduction in experiment times compared to conventional, many-pulsed, spectral decomposition sequences, while minimizing the effects of computational errors and experimental pulse imperfections.
\paragraph{}
The proposed method can be extended to any quantum system that suffers from decoherence due to spin-bath noise, whose spectrum $W(T,\alpha)$ can be calculated explicitly [similarly to Eq. \eqref{eq:exact}]. For example, a Gaussian noise with correlation function [equivalent to Eq. \eqref{eq:correlation}] of the form $b^2 \exp \left [ -\left(\frac{t}{\tau_c} \right)^2\right]$ will result in a similar expression to Eq. \eqref{eq:exact} with error functions replacing the internal exponential terms. Furthermore, additional effects of detuning arising from hyperfine splitting can be taken into account. For example, for an NV ensemble associated with $^{14}N$ atoms, a hyperfine structure with typical detunings $\Delta_1=2.2$ MHz and $\Delta_2=4.4$ MHz from the main resonance \cite{Jelezko2006} will lead to a modified coherence function $\tilde{W}(T,\alpha)=W(T,\alpha) \times OSC(\Delta_1,\Delta_2,T)$, incorporating a sum of oscillations at the detuning frequencies.
\section*{Acknowledgements}
This work has been supported in part by the Minerva ARCHES award, the CIFAR-Azrieli global scholars program, the Israel Science Foundation (grant No. 750/14), the Ministry of Science and Technology, Israel, and the CAMBR fellowship for Nanoscience and Nanotechnology.

\bibliography{nvbibliography}

\end{document}